\documentclass[prl,showpacs,showkeys,preprintnumbers,amsmath,amssymb,superscriptaddress,twocolumn]{revtex4-1}
\usepackage{graphicx}
\usepackage{amsfonts}
\usepackage{color}

\def\rp{p}

\def\xip{{\xi_{\rm per}}}

\def\tp{{t_{\rm per}}}

\def\rka{{\rho_{\rm ka}}}
\def\<{\langle}
\def\>{\rangle}
\begin{document}

\title{Dynamical Correlation Length and Relaxation Processes in a Glass Former}

\author{Raffaele Pastore}
\email{pastore@na.infn.it}
\affiliation{CNR--SPIN, Dip.to di Scienze Fisiche,
 Universit\'a di Napoli ``Federico II'', Naples,
Italy}

\author{Massimo Pica Ciamarra}
\affiliation{CNR--SPIN, Dip.to di Scienze Fisiche,
 Universit\'a di Napoli ``Federico II'', Naples, Italy}

\author{Antonio de Candia}
\affiliation{CNR--SPIN, Dip.to di Scienze Fisiche,
 Universit\'a di Napoli ``Federico II'', Naples,
Italy}

\author{Antonio Coniglio}
\affiliation{CNR--SPIN, Dip.to di Scienze Fisiche,
 Universit\'a di Napoli ``Federico II'', Naples,
Italy}

\date{Received: \today / Revised version: }

\begin{abstract}
We investigate the relaxation process and the dynamical heterogeneities
of the kinetically constrained Kob--Anderson lattice glass model, and show that these
are characterized by different timescales. The dynamics is well described
within the diffusing defect paradigm, which suggest to relate the
relaxation process to a reverse--percolation transition. This allows for
a geometrical interpretation of the relaxation process, and of the different timescales.
\end{abstract}


\maketitle
The hallmark of glass forming liquids, which is the rapid increase of the relaxation time
as the temperature decreases~\cite{Angell}, 
has been related to dynamical heterogeneities, growing spatio--temporal correlations in the dynamics.
Dynamical heterogeneities are predicted by theories of the glass transition such as
the mode--coupling theory, diffusing defects,
and the random first order theory~\cite{ToninelliE},
and have been observed in both experimental and numerical studies~\cite{Ediger}. 
These studies mostly focused on the dynamical susceptibility $\chi_4$,
whose maximum value $\chi_4^*$ estimates the number of
dynamically correlated particles. This maximum is expected to occur at a time
close to the relaxation time $\tau$, and to grow on approaching the transition
of structural arrest, as frequently observed. However, there exist systems where $\chi^*_4$
is found to decrease on approaching the transition~\cite{glotzer2000,fierro,drop},
as well as early studies suggesting that the time of maximal correlation between particles displacements
does not scale with the relaxation time $\tau$~\cite{Glotezer_times}.
Accordingly, the relation between the dynamical susceptibility and the relaxation process
remains elusive, and its clarification of great interest
as it would allow to contrast different theories of the glass transition.
Here we address this problem via a numerical study of the Kob--Anderson kinetically constrained
lattice gas model~\cite{KA}, where it is possible to obtain very accurate data
for the dynamical correlation length. We show that the relaxation process and the dynamical heterogeneities
are characterized by two different timescales, which implies that they are 
less tangled than expected.
We explain this feature in the diffusing defect picture, where we relate the relaxation process 
to a reverse percolation transition, and obtain a geometrical interpretation of the relaxation process and of the different timescales.
\begin{figure}[t!]
\begin{center}
\includegraphics*[scale=0.64]{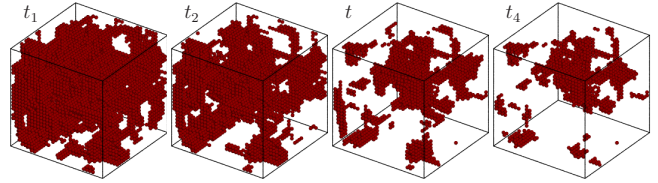}
\end{center}
\caption{\label{fig:plots}
(Color online)
Persistent particles in a numerical simulations of the Kob--Andersen model at $\rho = 0.85$, at times
$t_1 = 3.5~10^5$, $t_2 = 7.5~10^5$, $t_3 = 1.6~10^6$, and $t_4 = 2.1~10^6$.
}
\end{figure}

The Kob--Andersen lattice glass model~\cite{KA} is a kinetically constrained model~\cite{Ritort},
in which a particle is allowed to move in a near empty site if has less than $m = 4$ neighbors,
and if it will also have less than $m = 4$ neighbors after the move. Previous studies
have shown that this model reproduces many aspects of the dynamics of glass forming systems,
the slowing down of the dynamics on increasing the density suggesting the existence of a transition of
structural arrest at $\rho_{ka} = 0.881$~\cite{KA, FranzMulletParisi, Dawson},
even though it has been demonstrated that in the thermodynamic limit the transition of dynamical arrest only occurs
at $\rho = 1$~\cite{Toninelli}.
Here we investigate the relaxation process focusing on the time evolution
of the density of persistent particles $\rp(t)=\frac{1}V \sum_{i=1}^{V} n_{i}(t)$,
where $n_i(t) = 1(0)$ if site $i$ is (is not)
persistently occupied by a particle in the time interval $[0,t]$, and $\rho = N/V$ is the density.
$p(t)$ is related to the high wave vector limit of the intermediate
self scattering function~\cite{Chandler2006}.
As shown in Fig.~\ref{fig:plots}, as time proceeds the density of persistent particles decreases, and spatial correlations
between them emerge. These correlations are quantified by the dynamical susceptibility $\chi_4(t)$,
related to the fluctuations of $\rp$,
\begin{equation}
\label{eq:chip}
\chi_4(t)=\frac{V}\rho\left(\<\rp(t)^{2}\>-\<\rp(t)\>^{2}\right),
\end{equation}
and to the volume integral of the spatial correlation function between persistent particles at time $t$,
$\chi_4(t) = \frac{1}{\rho V}\sum_{i,j}^{V}g_4(r,t)$, where
\begin{equation}
\label{eq:gr}
g_4(r,t)=\<n_{i}(t)n_{j}(t)\>-\<n_{i}(t)\>\<n_{j}(t)\>,\,\,\, r = |i-j|.
\end{equation}
The spatial decay of $g_4(r,t)$ defines the dynamical correlation length $\xi(t)$.

\begin{figure}[t!]
\begin{center}
\includegraphics*[scale=0.33]{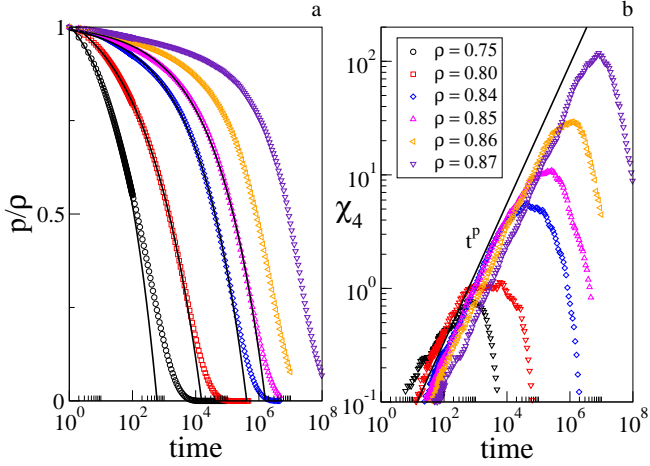}
\end{center}
\caption{\label{fig:review}
(Color online) Normalized density of persistent particles $\<\rp\>/\rho$ (panel a), and dynamical susceptibility $\chi_4$ (panel b),
for different values of the density, as indicated.
For $\rho \leq 0.85$, $\<\rp\>/\rho$ is well described by the von Schweidler law, $\<\rp\>/\rho = f_0- (t/\tau)^b$,
with $f_0 = 1$ and $b \simeq 0.3$. At short times, the dynamical susceptibility grows as $t^p$, with $p\simeq 0.61$.
}
\end{figure}

\begin{figure}[t!]
\begin{center}
\includegraphics*[scale=0.33]{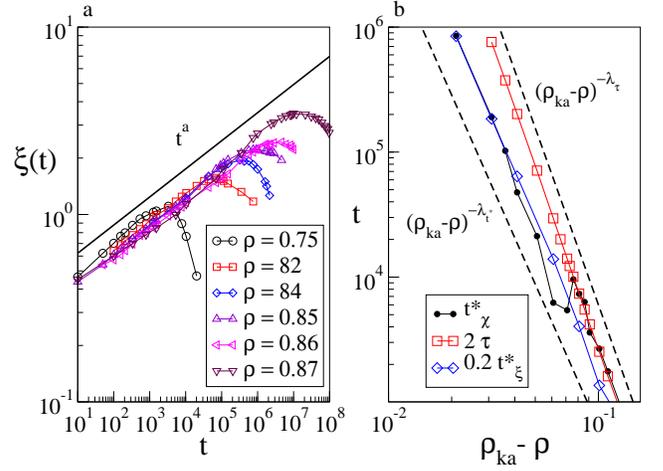}
\end{center}
\caption{\label{fig:xitau}
(Color online)
Panel a: dynamical correlation length for different values of the density.
Panel b: divergence of the relaxation time $\tau$, of the time where the correlation length acquires its maximum value $t^*_\xi$, and
of the time where the dynamical susceptibility acquires its maximum value, $t^*_\chi$. At low density, $t^*_\chi \propto \tau$,
while at high density $t^*_\chi \propto t^*_\xi$. Errors on $t^*_\xi$ and $t^*_\chi$ are of the order of $5\%$.
}
\end{figure}

\begin{figure}[t!]
\begin{center}
\includegraphics*[scale=0.33]{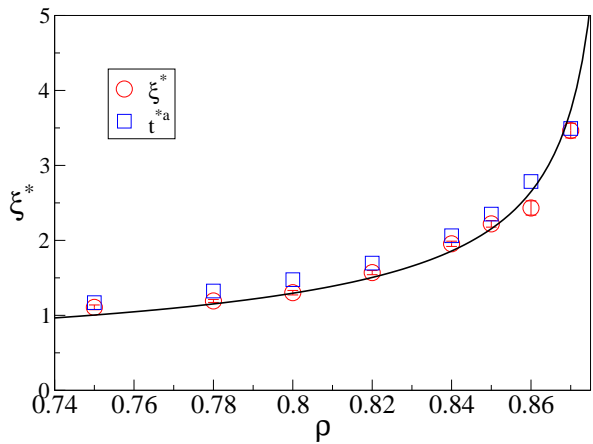}
\end{center}
\caption{\label{fig:lengths}
(Color online) Dynamical correlation length at $t = t^*$, and prediction of the diffusing defect picture,
$\xi^* \propto t^{*a} \propto \tau^q$, $q = a \lambda_\tau/\lambda_{t^*_\xi}$. The full line is a $(\rka-\rho)^{-\nu}$, $\nu \simeq 0.54$
(we finx $\rho_{ka} = 0.881$ as estimated from the divergence of the relaxation time).
}
\end{figure}

{\it Numerical results -- }
We start by shortly summarizing our study of the dynamics of the KA model~\cite{numerics}, which extends previous results
and allows to obtain new insights on the glassy dynamics. The values of the exponents
characterizing the dynamics are summarized in Table~\ref{table}.
Numerical results for $p(t)/\rho$ and $\chi_4(t)$ are shown in Fig.~\ref{fig:review}.
For $\rho \leq 0.85$,  the decay of $p(t)$ is well described by the von Schweidler law,
$\<\rp\>/\rho = f_0- (t/\tau)^b$, with $b \simeq 0.3$ and $f_0 \simeq 1$ in a large time window, while at higher densities $b$ increases
and $f_0$ decreases. The approach to the transition of structural arrest is marked by the increase of the relaxation time $\tau$,
we found to diverge as $\tau \propto \left(\rka-\rho\right)^{-\lambda_{\tau}}$,
as show in Fig.~\ref{fig:xitau}b, with  $\lambda_\tau = 4.7 \pm 0.1$ consistent with previous results~\cite{KA}.
The emergence of an increasingly heterogeneous dynamics is signaled by the dynamical susceptibility, initially growing as $\chi_4(t) \propto t^p$,
and then decreasing  after reaching its maximum value $\chi_4^*$ at a time $t^*_\chi$.
The dynamical correlation length, which has a similar behavior, is illustrated in Fig.~\ref{fig:xitau}a,
and is well described by
\begin{equation}
 \xi(t) \propto t^a \exp\left(-at/t^*_\xi\right).
\label{eq:xi}
\end{equation}
$\xi(t)$ grows as $t^a$ at short times, and then decreases after reaching its maximum value $\xi^*$ at time $t^*_\xi$,
we find to diverge as $t^*_\xi \propto \left(\rka-\rho\right)^{-\lambda_{t^*_\xi}}$, with $\lambda_{t^*_\xi} = 3.8 \pm 0.1$
The times $\tau$ and $t^*_\xi$ diverge with different exponents as $\rho$ approaches $\rka$.
This fact has rich consequences on the behavior of the susceptibility, we find to be well approximated by
\begin{equation}
\chi_4(t) \propto g(0,t) \xi(t)^{2-\eta} = p(t)(\rho-p(t)) \xi(t)^{2-\eta},
\label{eq:chi}
\end{equation}
with $\eta \simeq 0$~\cite{etazero}.
Indeed, at low densities $t^*_\xi \gg \tau$, and Eq.~\ref{eq:chi} predicts $t^*_\chi \propto \tau$,
while asymptotically $t^*_\xi \ll \tau$, and
the maximum of the susceptibility $\chi_4^*$ occurs at $t^*_\chi \propto t^*_\xi$.
Such a crossover in the behavior of $t^*_\chi$ is apparent in Fig.~\ref{fig:xitau}b.
In addition, when $t^*_\chi \propto \tau$, the maximum of the susceptibility
scales as $\chi_4^* \propto p(\tau)(\rho-p(\tau)) \tau^{2a} \propto \tau^{2a} \propto (\rka-\rho)^{-\gamma}$,
with $\gamma = 2a\lambda_\tau$, in agreement with our results.
Conversely, asymptotically $t^*_\chi \propto t^*_\xi$ and $\chi^*_4$ reflects the competition
between the amplitude, which vanishes as $p(t^*_\xi)(\rho-p(t^*_\xi)) \propto (\rka-\rho)^{-b(\lambda_{t^*_\xi}-\lambda_\tau)}$,
and the correlation length, which diverges as $\xi^* \propto {t^*_\xi}^a \propto (\rka-\rho)^{-\nu}$, with $\nu = a \lambda_{t^*_\xi} \simeq 0.54$
(Fig.~\ref{fig:lengths}). Consequently $\chi_4^* \propto (\rka-\rho)^{-q}$, with
$q =b(\lambda_{t^*_\xi}-\lambda_\tau) + 2a\lambda_{t^*_\xi} \simeq 0.9$.
Note that in other systems, where such decoupling between $\tau$ and $t^*_\xi$ may occur with a negative $q$ value, 
the susceptibility would  decrease on approaching the transition of structural arrest, as observed in some
experimental~\cite{drop} and numerical~\cite{glotzer2000,fierro} studies.

\begin{figure}[t!]
\begin{center}
\includegraphics*[scale=0.33]{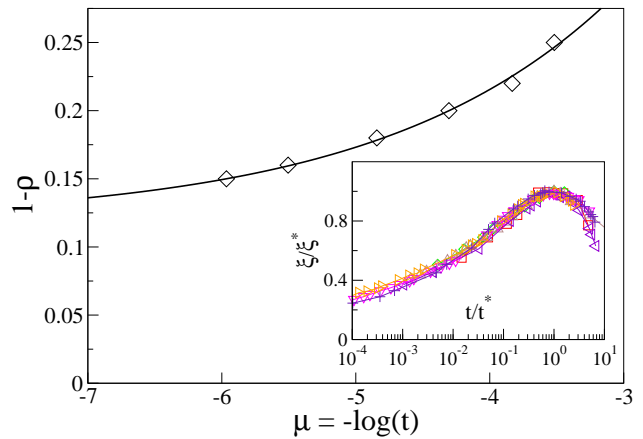}
\end{center}
\caption{\label{fig:widom}
(Color online)
Main panel: Widom line in the density, chemical potential plane. Circles indicates the time $t^*_\xi$
where the correlation length reaches its maximum value, at each value of the density. The continuous line corresponds
to $(1-\rka) \propto (t/t^*)^{a/\nu}$, and suggests that the system approaches a critical point at $\mu = -\infty$, and $\rho = \rka$.
Inset: scaling of the dynamical susceptibility for $7$ values of the density, in the range $0.78$--$0.87$.
}
\end{figure}

The presence of a growing correlation length suggests that the system is approaching a critical point as the density increases.
This scenario is conveniently described interpreting $\mu = -\log(t)$ as a chemical potential for the persistent particles,
considering that the density of persistent particles monotonically decreases as time advances.
The line where the correlation length reaches its maximum value in the $\mu$--$\rho$ plane
can therefore be interpreted as a Widom line, which in a second order transition ends at the critical point.
The results of Fig.~\ref{fig:widom} suggest the presence of a critical point located at $\rho = \rka$ and $\mu = -\infty$,
where the correlation length diverges. The Widom line will actually eventually bend,
and end at $\rho = 1$ where the transition is known to occur in the the thermodynamic limit.
Such an approach may open the way to a renormalization group treatment of the glass transition.

\begin{table}[b!]
\begin{tabular}{|c|c|c|}
\hline exponent  & measure & prediction \\
\hline  $p(t) = \rho\left(1-(t/\tau)^b\right)$   & $b = 0.3$ & -- \\
\hline  $\xi(t) \propto t^a$  &  $a = 0.156$ & $a = b/2 = 0.15$ \\
\hline  $\chi_4(t) \propto t^p$  & $p = 0.6$ & $p = 2b = 0.6$ \\
\hline  $\tau \propto \left(\rka-\rho\right)^{-\lambda_{\tau}}$  & $\lambda_{\tau} = 4.7$ & -- \\
\hline  $t^*_\xi \propto \left(\rka-\rho\right)^{-\lambda_{t^*_\xi}}$  & $\lambda_{t^*_\xi} = 3.8$ & -- \\
\hline  $\xi^* \propto \left(\rka-\rho\right)^{-\nu}$  & $\nu = 0.54$ & $\nu = a\lambda_{t^*_\xi} = 0.57$  \\
\hline  $\chi_4^* \propto \left(\rka-\rho\right)^{-\gamma}$  & $\gamma=1.43$ & $\gamma = 2a\lambda_\tau = 1.41$ ($t^*_\xi \gg \tau)$ \\
\hline
\end{tabular}
\caption{\label{table}Exponents characterizing the slow dynamics of the KA model, and their relations according to the diffusing
defects picture. The fractal dimension is $d_f = b/a$. The dynamics is characterized by three exponents, $b$, $\lambda_\tau$ and
$\lambda_{t^*_\xi}$.}
\end{table}

{\it Diffusing defects -- }
The results described so far are rationalized in the diffusion defects paradigm~\cite{ToninelliE,reviewBB,Shlesinger}, where
the relaxation is ascribed to the presence of possibly extended diffusing defects, with density $\rho_d$.
The number of distinct sites visited by a defect grows as $n_v(t) \propto t^{b}$, while its mean square
displacement grow as $t^{2a}$, $d_f = b/a$ being the defect fractal dimension. At short times, before defects interact,
the persistence decays as $1-p(t)/\rho \propto \rho_d n_v(t) \propto \rho_d t^b$, and therefore this picture
reproduces the von Schweidler law, and relates the density of defects $\rho_d$ with relaxation time, $\rho_d \propto \tau^{-b}$.
The correlation length is expected to grow as $t^a$ as long as different defects do not interact,
since only sites visited by the same defect are correlated. Due to their sub--diffusive nature, we expect defects
to behave as random walkers characterized by a fat--tail waiting time distribution~\cite{Yuste}, which does not affect
their fractal dimensions. The value of the fractal dimension is also largely unaffected by the possible presence
of spatial correlations~\cite{attractiveRW}.
Accordingly, the diffusing defect picture predicts $b/a = d_f = 2$, in agreement with the numerical findings.
This picture also predicts that the susceptibility grows as the square of the number of sites visited by each defect~\cite{ToninelliE},
$\chi_4(t) \propto \rho_b n_v(t)^2 \propto t^{2b}$, which allows to correctly estimate $p = 2b$.

\begin{figure}[t!]
\begin{center}
\includegraphics*[scale=0.33]{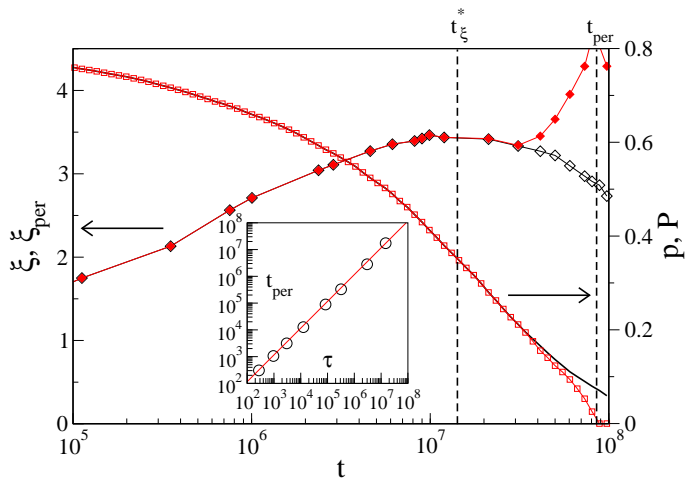}
\end{center}
\caption{\label{fig:rho85}
(Color online) Percolation transition at $\rho = 0.87$. Left axis:
dynamical correlation length $\xi$ (empty diamonds) and percolation correlation
length $\xip$ (full diamonds). Right axis: density of persistent particles $\<\rp\>$ (full line)
and strength of the percolating cluster $\<P\>$ (squares).
The vertical dashed lines mark $t^*_\xi$ and $\tp$, which is proportional to $\tau$ (inset).
}
\end{figure}

{\it Reverse percolation -- }
As time advances defects induce a reverse percolation transition of persistent particles, clearly visible in Fig.~\ref{fig:plots},
which is similar to the gradual destruction of a polymer gel through diffusing enzymes~\cite{Abete,Sidoravicius}.
Since the absence of a percolating cluster of persistent particles leads to the lost of
mechanical rigidity on all timescales, this transition is related to the relaxation process.
Indeed, the study of the density of the percolating cluster $P$,
reveals that the percolation time scales with the relaxation time, as shown in Fig.~\ref{fig:rho85}.
The figure also reveals that $P$ equals $p$ up to large times, which implies that 
the density of finite clusters $p$-$P$ is negligible during most of the relaxation process.
At large times, finite clusters appear and have a broad size distribution, and consequently
different relaxation timescales, thus explaining the crossover in decay of $p(t)$, 
which is first described by a power-law, and then by a stretched exponential.
Fig.~\ref{fig:rho85} also shows that the dynamical correlation length coincides with the percolative length
extracted from the pair connected correlation function~\cite{Coniglio}, as long as finite clusters are negligible.
The percolative length is affected by the two timescales characterizing
the glassy dynamics, the time $t = t^*_\xi$ where the dynamical length reaches its maximum value, and the percolating time related
to the relaxation of the system. At high densities, this makes $\xip$ non monotonic, as in Fig.~\ref{fig:rho85}.

{\it Future directions -- } Our results suggest that in the KA
model the relaxation dynamics and the dynamical heterogeneities
are characterized by different timescales, $\tau$ and $t^*_\xi$.
This may be explained considering that $\tau$ occurs when a given
large fraction of all sites has relaxed, while $t^*_\xi$ occurs
when the correlations between the persistent particles decreases.
In principle, these correlations may decrease before a large
fraction of all sites has relaxed. Since $\tau \propto t^*_\xi$
when defects behave as perfect random walkers~\cite{ToninelliE},
one may speculate that the decoupling  between the two timescales
may be due to a more complex nature of defects, such as the non
conservation of defects characterized by birth and death rate 
with a constant average number or the presence of heterogeneous
defects. Future plans include the study of the dynamical
correlation length in off-lattice models of glass forming liquids,
to verify the possible existence of different timescales.


\begin{thebibliography}{30}
\bibitem{numerics}
We have performed Monte Carlo simulations of the standard Kob--Andersen (KA) model
in three dimensions~\cite{KA}, for $L = 30$, performing up to $10^8$ sweeps. For each value of
the density, we have averaged our results over up to $10^4$ runs.
\end{thebibliography}
\end{document}